\documentclass[12pt]{article}
\usepackage{pic02}
\usepackage{hyperref}
\usepackage{url}
\usepackage{psfig}
\usepackage[numbers]{natbib}

\def \apj{Astrophys. J.}
\def \apjl{Astrophys. J.}

\def \aap{Astron. Astrophys.}

\def \ca{Comments on Astrophys.}

\def \mnras{Mon. Not. Roy. Soc.}
\def \nat{Nature}
\def \npb{Nucl. Phys. B}

\def \pra{Phys. Rev. A}

\def \prd{Phys. Rev. D}
\def \prl{Phys. Rev. Lett.}

\def \rmp{Rev. Mod. Phys.}

\def \da{\Delta\alpha/\alpha}

\begin{document}

\title{\bf TIME EVOLUTION OF THE FINE STRUCTURE CONSTANT}
\author{\underline{M. T. Murphy}, J. K. Webb, V. V. Flambaum,
S. J. Curran\\ {\em School of Physics, University of New South Wales,
Sydney N.S.W. 2052, Australia}}

\maketitle

%
%
%
%
%
%
\vspace{4.5cm}
%

\baselineskip=14.5pt
\begin{abstract}
We present a short review of the current quasar (QSO) absorption line
constraints on possible variation of the fine structure constant, $\alpha
\equiv e^2/\hbar c$. Particular attention is paid to recent optical
Keck/HIRES spectra of 49 absorption systems which indicate a smaller
$\alpha$ in the past \cite{MurphyM_01a,WebbJ_01a}. Here we present new
preliminary results from 128 absorption systems: $\da = (-0.57 \pm
0.10)\times 10^{-5}$ over the redshift range $0.2 < z < 3.7$, in agreement
with the previous results. Known potential systematic errors cannot explain
these results. We compare them with strong `local' constraints and discuss
other (radio and millimetre-wave) QSO absorption line constraints on
variations in $\alpha^2g_p$ and $\alpha^2g_pm_e/m_p$ ($g_p$ is the proton
$g$-factor and $m_e/m_p$ is the electron/proton mass ratio). Finally, we
discuss future efforts to rule out or confirm the current 5.7\,$\sigma$
optical detection.
\end{abstract}
\newpage

\baselineskip=17pt

\section{Introduction}\label{sec:intro}

The assumption that the constants of Nature remain constant in spacetime
should be experimentally tested \cite{BekensteinJ_79a}. Strong motivation
for {\it varying} constants comes from modern unified theories
\cite{MarcianoW_84a,BarrowJ_87a,DamourT_94a}. Here we review the QSO
absorption line constraints on possible variation of the electromagnetic
coupling constant, $\alpha$. Our most recent published results are
summarized in reference \cite{WebbJ_01a} and in Section \ref{sec:results}
we present new preliminary results from a significantly extended optical
sample.

\section{QSO absorption systems and the alkali doublet
  method}\label{sec:AD}

For small variations in $\alpha$, the relative wavelength separation
between the transitions of an alkali doublet (AD) is proportional to
$\alpha$. Savedoff \cite{SavedoffM_56a} first utilized this to constrain
possible variations in $\alpha$ from AD separations seen in galaxy emission
spectra. The advantage of this technique is the large look-back times
inherent in such cosmological observations ($\sim 10{\rm
\,Gyr}$). Absorption lines produced by intervening clouds along the line of
sight to QSOs are substantially narrower than intrinsic emission lines and
so yield tighter limits on $\alpha$-variation \cite{WolfeA_76a}.

Spectrographs on 8--10-m optical telescopes can record high resolution
(${\rm FWHM} \sim 7{\rm \,kms}^{-1}$), high signal-to-noise (${\rm S/N}
\sim 30{\rm \,per~pixel}$) spectra of high redshift QSOs over most of the
optical range (i.e. 3000--8000\,\AA) in several $\sim 1\,{\rm hr}$
exposures. Fig. 1 shows an example QSO spectrum with a C{\sc \,iv} AD. Many
velocity components of the absorption system are clearly resolved.

\begin{figure}[t]
\centerline{\psfig{file=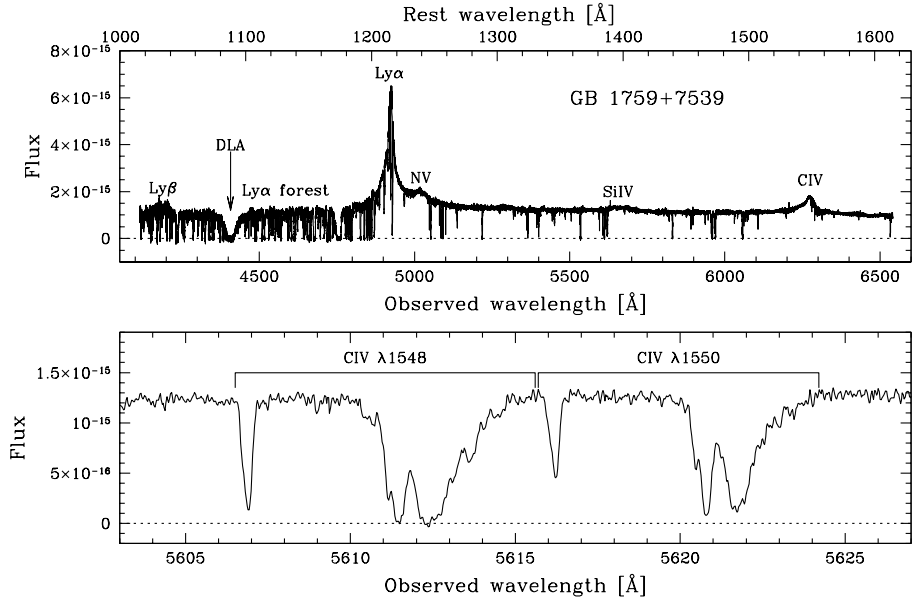,width=120mm}}
\caption{\it Keck/HIRES spectrum of QSO GB 1759+7539
  \cite{OutramP_99a}. The full spectrum (upper panel) shows several
  emission lines intrinsic to the QSO (Ly-$\alpha$, Ly-$\beta$, N{\sc
  \,iv}, Si{\sc \,iv}, C{\sc \,iv}). The damped Ly-$\alpha$ system (DLA) at
  $z_{\rm abs}=2.6253$ gives rise to heavy element absorption lines in the
  red portion of the spectrum. The lower panel details a small region
  containing a C{\sc \,iv} alkali doublet. The separation between
  corresponding velocity components in the two transitions is proportional
  to $\alpha$ for $\da \ll 1$.}
\end{figure}

Varshalovich et al. \cite{VarshalovichD_00a} have recently used spectra of
16 Si{\sc \,iv} ADs with a mean redshift $\left<z_{\rm abs}\right> = 2.6$
to obtain a value for the fractional difference between $\alpha$ in the
laboratory and in the QSO spectra, $\da \equiv (\alpha_z -
\alpha_0)/\alpha_0 = (-4.6 \pm 4.3_{\rm stat} \pm 1.4_{\rm sys})\times
10^{-5}$. The systematic error term arose from uncertainties in the
laboratory wavelengths of the Si{\sc \,iv} transitions: the astronomical
spectra were of comparable quality to UV laboratory spectra. Significant
improvements in the laboratory wavelengths \cite{GriesmannU_00a} reduce
this systematic error to $0.2 \times 10^{-5}~(1\,\sigma)$. We have analyzed
21 high quality Si{\sc \,iv} doublets observed with the HIRES spectrograph
on the Keck I 10-m telescope in Hawaii, finding \cite{MurphyM_01c}
\begin{equation}\label{eq:SiIV}
\da = (-0.5 \pm 1.3_{\rm stat})\times 10^{-5}
\end{equation}
at $\left<z_{\rm abs}\right> = 2.8$. The factor of 3 improvement in
precision is due to the high spectral resolution of the HIRES data. This is
currently the strongest constraint on $\da$ from the AD method.

\section{The many-multiplet method}\label{sec:MM}

The AD method is simple, but inefficient. The $s$ ground state is most
sensitive to changes in $\alpha$ (i.e. it has the largest relativistic
corrections) but is common to both transitions (Fig. 2a). A more sensitive
method is to compare transitions from different multiplets and/or atoms,
allowing the ground states to constrain $\da$ (Fig. 2b). This is the many
multiplet (MM) introduced in \cite{DzubaV_99a,WebbJ_99a}.

\begin{figure}[t]
\centerline{\psfig{file=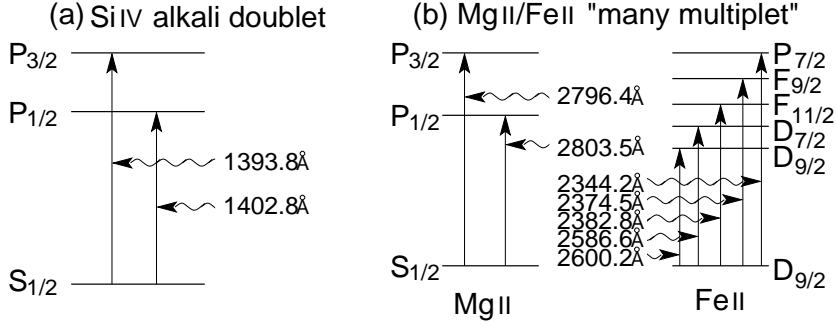,width=110mm}}
\caption{\it (a) The AD method is not sensitive to the maximal relativistic
  corrections in the common $S$ ground state. (b) Comparison of different
  ions increases sensitivity to $\da$, increases statistics and decreases
  systematic errors.}
\end{figure}

To illustrate the MM method, consider the following semi-empirical equation
for the relativistic correction, $\Delta$, for a transition from the ground
state with total angular momentum, $j$:
\begin{equation}\label{eq:delta}
\Delta \propto (Z\alpha)^2\left[\frac{1}{j+1/2} - C\right]\,,
\end{equation}
where $Z$ is nuclear charge and many-body effects are described by $C \sim
0.6$. To obtain strong constraints on $\da$ one can (a) compare transitions
of light ($Z \sim 10$) atoms/ions with those of heavy ($Z \sim 30$) ones
and/or (b) compare $s$--$p$ and $d$--$p$ transitions of heavy elements. For
the latter, the relativistic corrections will be of opposite sign which
further increases sensitivity to $\alpha$-variation and strengthens the MM
method against systematic errors in the QSO spectra (see Section
\ref{sec:syserr}).

More formally, we may write the following equation for the rest-frequency,
$\omega_z$, of {\it any} transition observed in the QSO spectra at a
redshift $z$:
\begin{equation}\label{eq:omega}
\omega_z = \omega_0 + q\left[\left(\frac{\alpha_z}{\alpha}\right)^2-1\right]\,,
\end{equation}
where $\omega_0$ is the frequency measured in the laboratory on Earth (we
omit higher order terms here for simplicity). Laboratory measurements
\cite{PickeringJ_98a,PickeringJ_00a,GriesmannU_00a} of $\omega_0$ for many
transitions commonly observed in QSO spectra now allow a precision of $\da
\sim 10^{-7}$ to be achieved. The $q$ coefficient contains all the
relativistic corrections and measures the sensitivity of each transition
frequency to changes in $\alpha$. These have been calculated in
\cite{DzubaV_99a,DzubaV_99b,DzubaV_01a,DzubaV_02a} to $<10\%$ precision
using the Dirac-Hartree-Fock approximation and many-body perturbation
theory. Note that the form of Eq. \ref{eq:omega} ensures one cannot infer a
non-zero $\da$ due to errors in the $q$ coefficients.

Fig. 3 shows the distribution of $q$ coefficients in (rest) wavelength
space. Our sample conveniently divides into low- and high-$z$ subsamples
with very different properties. Note the simple arrangement for the low-$z$
Mg/Fe{\sc \,ii} systems: the Mg transitions are used as anchors against
which the large, positive shifts in the Fe{\sc \,ii} transitions can be
measured. Compare this with the complex arrangement for the high-$z$
systems: low-order distortions to the wavelength scale will have a varied
and complex affect on $\da$ depending on which transitions are fitted in a
given absorption system. In general, the complexity at high-$z$ will yield
more robust values of $\da$.

\begin{figure}[t]
\centerline{\psfig{file=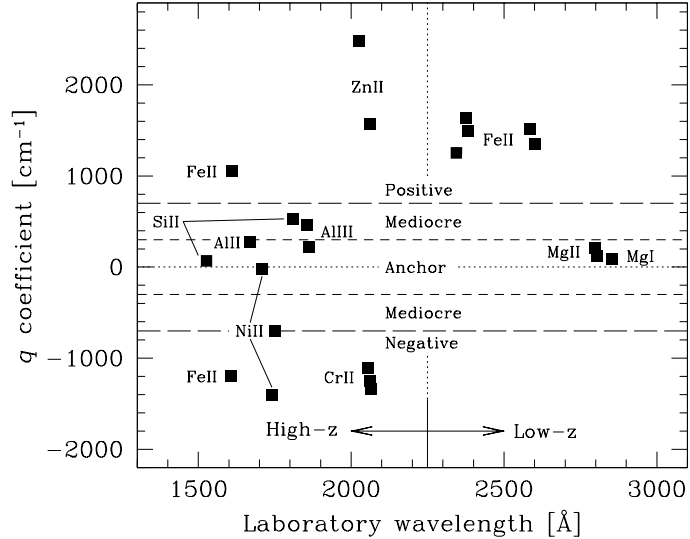,width=90mm}}
\caption{\it Distribution of $q$ coefficients for the transitions used in
  the MM method. For the low-$z$ Mg/Fe systems, a compression of the
  spectrum can mimic $\da<0$. However, the complex arrangement at high-$z$
  indicates resistance to such systematics.}
\end{figure}

\section{Recent results}\label{sec:results}

For each absorption system we fit multiple velocity component Voigt
profiles to all available (typically $\sim$5) MM transitions. We minimize
$\chi^2$ for all velocity components {\it simultaneously} to obtain the
best fitting value of $\da$. The 1\,$\sigma$ error is derived from the
diagonal terms of the final parameter covariance matrix. Monte Carlo
simulations demonstrate the reliability of both $\da$ and the errors.

The MM method was first applied to 30 low-$z$ Mg/Fe systems and provided a
tentative non-zero $\da$ \cite{WebbJ_99a}. In \cite{MurphyM_01a} we
extended the sample of \cite{WebbJ_99a} to 49 absorption systems, finding
4.1\,$\sigma$ evidence for a smaller $\alpha$ in the redshift range $0.5 <
z_{\rm abs} < 3.5$. We have now increased our sample to 128 absorption
systems, all observed with Keck/HIRES. Our new preliminary weighted mean is
\begin{equation}
\da = (-0.57 \pm 0.10)\times 10^{-5}
\end{equation}
for $0.2 < z_{\rm abs} < 3.7$, i.e. 5.7\,$\sigma$ statistical evidence for
a smaller $\alpha$ in high redshift absorption systems. We plot $\da$
versus $z_{\rm abs}$ in Fig. 4. Note the overall internal consistency of
the results. Fig. 5 suggests possible evolution of $\alpha$ with
cosmological time, although see Section \ref{sec:limits} for further
discussion and caveats of fixing $\da = 0$ at $z=0$.

\begin{figure}[t]
\centerline{\psfig{file=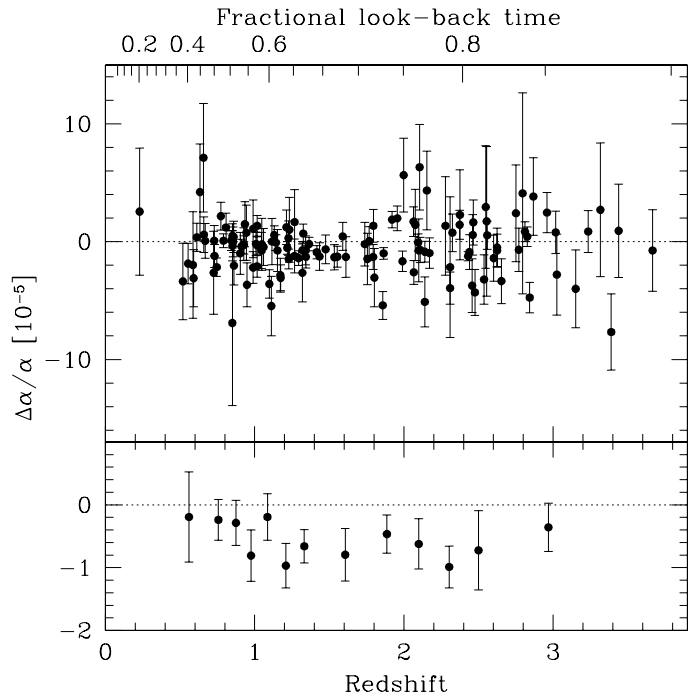,width=90mm}}
\caption{\it Distribution of $\da$ over absorption redshift. The upper
  panel shows $\da$ for 128 absorption systems with 1\,$\sigma$ errors. We
  bin $\da$ in the lower panel, presenting the weighted mean $\da$ and
  1\,$\sigma$ error at the mean redshift for each bin.}
\end{figure}

\begin{figure}[t]
\centerline{\psfig{file=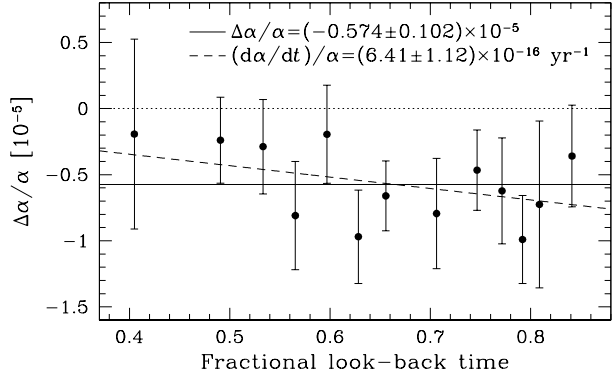,width=80mm}}
\caption{\it Temporal variation in $\alpha$. The points are the binned
values of $\da$ from Fig. 4, the solid line is the weighted mean and the
dashed line is a fit to the raw (i.e. unbinned) data fixed to $\da = 0$ at
$z=0$. A $\chi^2$ analysis indicates that an evolving $\da$ is preferred.}
\end{figure}

\section{Systematic errors?}\label{sec:syserr}

The statistical error in this result is now small: we do detect line shifts
in the QSO spectra. But are the line shifts due to systematic errors or
really due to varying $\alpha$? We have thoroughly searched for possible
systematic errors in our previous results \cite{MurphyM_01b}, finding none
which provide an alternative interpretation of the data. We have extended
this search to the new data in Fig. 4 with similar results. Currently, our
two largest sources of possible systematic error are:
\begin{enumerate}
\item Atmospheric dispersion effects: Before 1996 Keck/HIRES had no image
  rotator and so the effects of atmospheric dispersion on the wavelength
  scale could not be avoided. Effective compression of the spectra may
  result, possibly mimicking a negative $\da$ at low-$z$. 77 of our 128
  absorption systems could have been affected. However, we find {\it no
  evidence for these effects}: the ``affected'' and ``unaffected''
  subsamples yield the same $\da$. Nevertheless, we modeled the potential
  effect and correct the 77 affected systems in Fig. 6 (top panel). This
  correction reduces the significance of the low-$z$ points but increases
  the significance of those at high-$z$, enhancing the apparent trend in
  $\da$ with $z$.

\item Isotopic ratio evolution: We fit Mg and Si absorption lines with
  terrestrial values of the isotopic ratios. If the isotopic abundances in
  the absorption clouds are different to the terrestrial values then we may
  introduce artificial line shifts, potentially leading to $\da \neq
  0$. Galactic observations \cite{GayP_00a} and theoretical models
  \cite{TimmesF_96a} strongly suggest that only the $^{24}$Mg and $^{28}$Si
  isotopes will exist in the absorption clouds with significant
  abundances. The middle panel of Fig. 6 shows that the low-$z$ points
  become {\it more} significant when we fit only these isotopes to our
  data.
\end{enumerate}

\begin{figure}[t]
\centerline{\psfig{file=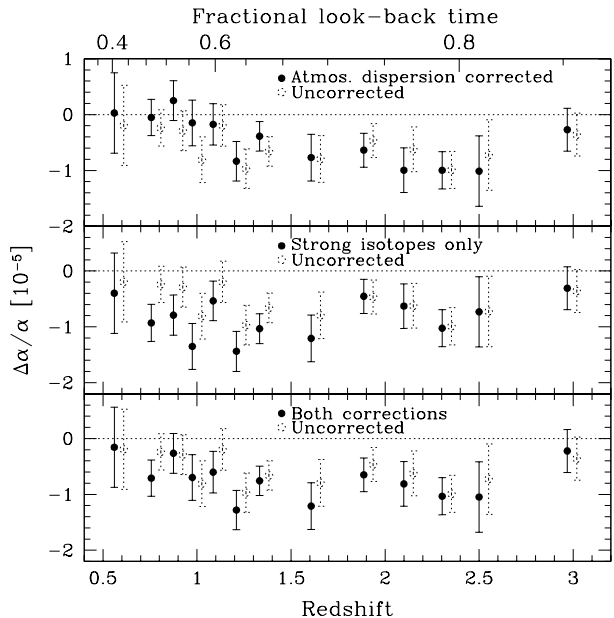,width=80mm}}
\caption{\it Summary of systematic effects. The top two panels compare the
  uncorrected results in Fig. 4 with those corrected for our two most
  important potential systematic effects: atmospheric dispersion and
  isotopic ratio evolution. The lower panel combines the two corrections.}
\end{figure}

The above two effects cannot explain our results. Indeed, applying both
corrections (lower panel of Fig. 6), yields a more significant result.

\section{Other QSO absorption line methods}\label{sec:radio}

Comparing absorption lines of the hydrogen hyperfine (21-cm) transition and
millimetre-wave molecular rotational transitions offers an order of
magnitude gain in precision over the MM method (per absorption system). The
21-cm/mm frequency ratio is $\propto \alpha^2g_p$ \cite{DrinkwaterM_98a}:
$\da \neq 0$ manifests itself as a difference between the 21-cm and mm
absorption redshifts. A similar difference may arise between 21-cm and
optical absorption lines, in this case constraining $\alpha^2g_pm_e/m_p$
\cite{CowieL_95a}. Here, $g_p$ is the proton $g$-factor and $m_e/m_p$ is
the electron-proton mass ratio.

However, systematic errors in these techniques are more difficult to
quantify than for the optical MM method. Since the radio and mm continuum
emission come from separate regions of the background QSO, the 21-cm and mm
absorption may occur along slightly different sight-lines. Thus, a
statistical sample of such measurements is required. Unfortunately, due to
the paucity of known absorption systems, only two 21-cm/mm comparisons
\cite{CarilliC_00a,MurphyM_01d} and one 21-cm/optical comparison
\cite{CowieL_95a} presently exist (Fig. 7).

\section{Other limits on varying $\alpha$}\label{sec:limits}

We summarize the strongest current constraints on $\alpha$-variation in
Fig. 7. For brevity, we do not discuss the reliability of the `local'
constraints, and refer the reader to \cite{UzanJ_02a} for a review.

Instead we focus on a comparison of the local and cosmological
constraints. Despite the tight limits on $\da$ from laboratory atomic
clocks, the Oklo phenomenon and meteoritic $\beta$-decay, a simple
non-linear evolution of $\alpha$ with time can explain all results
simultaneously. Moreover, we emphasize that it is dangerous to compare
local and cosmological limits without a better understanding of possible
spatial variations in $\alpha$ \cite{BekensteinJ_79a,BarrowJ_01a}. For
example, absorption spectroscopy of $z_{\rm abs} \approx 0$ absorption
clouds in our Galaxy may not yield $\da=0$ (cf. Fig 5). Even comparing the
different QSO absorption constraints is difficult since the MM and AD
methods constrain $\alpha$ whereas the 21-cm/mm and 21-cm/optical methods
constrain $\alpha^2g_p$ and $\alpha^2g_pm_e/m_p$ respectively.

\begin{figure}[t]
\centerline{\psfig{file=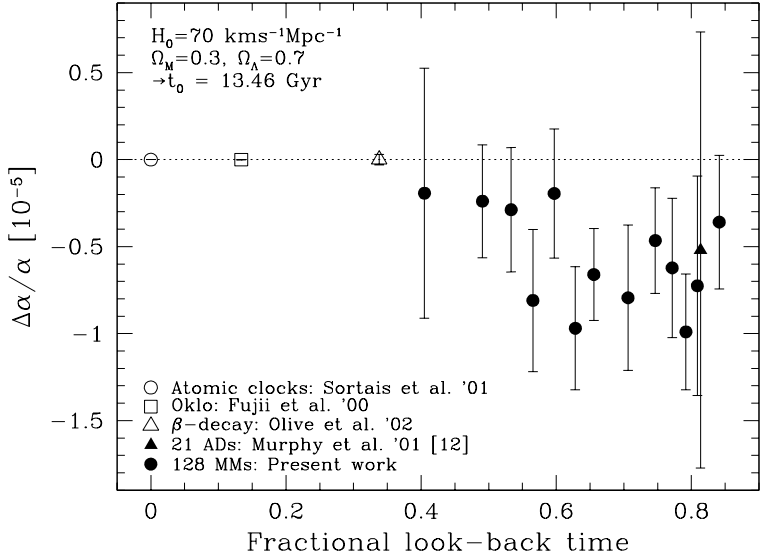,width=100mm}}
\caption{\it Comparison of the strongest current constraints on $\da$ from
  `local' tests (open symbols) and QSO absorption lines (solid symbols).}
\end{figure}

\section{The future}\label{sec:future}

Although the results in Fig. 4 would have tremendous theoretical
implications, confirming or refuting them is an observational issue. We are
taking two main steps to check our recent results:
\begin{enumerate}
\item Independent optical data. The greatest present concern is that only
  one instrument has been used for all our observations. QSO spectra of
  similar quality to the Keck/HIRES data are now becoming available. Data
  from other telescopes/instruments (e.g. VLT/UVES) will provide an
  important check on our results.

\item Further 21-cm/mm/optical comparisons. We are carrying out
  observations aimed at identifying new H{\sc \,i} 21-cm and mm-band
  molecular rotational absorption systems
  (e.g.~\cite{CurranS_02b}). Obtaining a statistical sample of 21-cm/mm and
  21-cm/optical comparison is vital for negating the line-of-sight velocity
  differences discussed above.
\end{enumerate}

If step (1) confirms our Keck/HIRES results then step (2) will be a crucial
check with entirely different systematic errors.

\section*{Acknowledgments}
We acknowledge the continued support of Chris Churchill, Jason Prochaska
and Arthur Wolfe who provided the previous Keck/HIRES datasets. We are
indebted to Wallace Sargent for the new Keck/HIRES dataset. We are grateful
to the John Templeton Foundation for supporting this work. MTM thanks the
PIC organizing committee for organizing financial support to speak at this
conference.


\end{document}